\documentclass[aps,prb,preprint,amsmath,longbibliography]{revtex4-2}
\usepackage{bm}
\usepackage{graphicx}
\usepackage{color}
\usepackage{amsmath}
\usepackage{amsfonts}
\usepackage{amssymb}
\usepackage{dcolumn}
\usepackage{hyperref}
\hypersetup{colorlinks=true, citecolor=blue, linkcolor=blue,urlcolor=blue
}

\begin{document}

\title{Lower (negative) bounds on the static electric susceptibility of non-equilibrium cubic crystals}

\author{R.~Dutta}
\affiliation{School of Electronic Engineering and Computer Science, Queen Mary University of London, Mile End Road, London E1~4NS, United Kingdom}
\author{P.~D.~Wurzner}
\affiliation{School of Electronic Engineering and Computer Science, Queen Mary University of London, Mile End Road, London E1~4NS, United Kingdom}
\author{F.~Castles}
\email{f.castles@qmul.ac.uk}
\affiliation{School of Electronic Engineering and Computer Science, Queen Mary University of London, Mile End Road, London E1~4NS, United Kingdom}

\begin{abstract}

For a long time, conventional wisdom held that the static electric susceptibility of all materials must be positive.  However, it is now believed that negative static electric susceptibility is likely to be possible in certain materials that are not in thermodynamic equilibrium (though conclusive experimental evidence of this remains to be reported).  Negative static electric susceptibility represents a qualitatively new parameter range of an important material property and it is thought that it may open up the possibility of new physical effects and technological capabilities.  With the recent introduction of active metamaterials composed of meta-atoms that appear to be capable, in principle, of exhibiting negative static polarizabilities of arbitrarily large magnitudes, it seems timely to investigate how the mutual interaction between dipolarizable entities with negative static polarizability affects the stability of such systems, and, in particular, to determine the lower bound on the static electric susceptibility that may be achieved in non-equilibrium materials.  Here we use a classical, microscopic model of point-like dipolarizable entities (a model that is standard in the case of positive polarizability) and investigate its behavior for simple cubic (sc), body-centered cubic (bcc), and face-centered cubic (fcc) crystals with one entity per primitive cell when the static polarizability of the entities is negative and the mutual interaction between the entities is taken into account.  We find that the static electric susceptibility is bounded below due to an instability towards self-polarization but the lower permissible bound is negative definite in each case, i.e., the concept of negative static electric susceptibility remains robust, according to the model, when mutual interactions are taken into account.  The usual Clausius-Mossotti relation between the static polarizability and the static electric susceptibility remains valid in the case of negative parameters, but only down to the lower permissible bound; the value of the bound depends on the crystal structure and is always unrelated to the asymptote of the Clausius-Mossotti curve.  The lower permissible bounds of the static electric susceptibility are found to be $-0.906$ for sc, and $-1.00$ for bcc and fcc.  These results confirm that, although the magnitude of the static electric susceptibility does not diverge in the negative case (as it can in the positive case), the magnitudes attainable in the negative case for condensed media may, nevertheless, be many orders of magnitude greater than those predicted previously for inverted vapors and gases.  This is a promising result in relation to the development of potential new technologies that exploit the phenomenon.

\end{abstract}


\maketitle

\section{Introduction}

The discovery and development of materials with new or enhanced properties is an important driver of technological advance.  In the area of electrostatics and low-frequency electromagnetics, a key material property is static electric susceptibility $\chi^{(0)}$ (or, equivalently, static relative permittivity $\varepsilon^{(0)}$, related via $\varepsilon^{(0)}=\chi^{(0)}+1$).  The value of $\chi^{(0)}$ for a material can be crucial to the utility of the material in certain device applications; for example, its value is of crucial importance in capacitor dielectrics, where a large (positive) value is often desired to facilitate device miniaturization \cite{hippel_1954,reynolds_2004}.  Since the first measurements of Cavendish \cite{cavendish_1921} and Faraday \cite{faraday_ERE_11_series_phil_trans}, it was found that the presence of a material between two conductors always increased the static mutual capacitance of the conductors above the capacitance observed when no material was present, i.e., it became established empirically that $\varepsilon^{(0)}>1$ and, correspondingly, $\chi^{(0)}>0$ for all materials \footnote{For completeness, we may note that, following Faraday, some relatively early researchers considered air, rather than vacuum, as the reference medium, even when measurement techniques became sufficiently precise to distinguish between the values obtained for various gases.  Therefore, values of $\varepsilon^{(0)}$ (or the specific inductive capacity, as it was called initially) slightly less than unity have been reported for gases such as Hydrogen: see, e.g., Ref.~\cite{thomson_jj_1921}.  However, since it has been accepted subsequently that the more-appropriate reference is vacuum, the values of $\varepsilon^{(0)}$ reported experimentally are greater than unity for all materials (but may be less than that of air for certain gases): see, e.g., Ref.~\cite{kaye_laby_1973}}.  Although early researchers kept an open mind regarding whether, as-yet untested, materials may nevertheless be found subsequently to exhibit $\varepsilon^{(0)}$ values less than unity (see, e.g., Ref.~\cite{maxwell_1873_p_65}), a theoretical argument was developed later that appeared to rule out this possibility; in \S14 of Ref.~\cite{ll8}, Landau \textit{et al}. conclude that ``the permittivity of all bodies exceeds unity, and the dielectric susceptibility ... is therefore positive''.  (Note that Landau \textit{et al.} are referring here to the \textit{static} relative permittivity and electric susceptibility).  It has since been noted, however, that Landau \textit{et al}.'s argument assumes that the bodies in question are in thermodynamic equilibrium, and, therefore, the conclusion does not necessarily hold for media that are not in thermodynamic equilibrium \cite{sanders_1986,sanders_1988,chiao_1994,chiao_1995_1,chiao_1995_2,chiao_1995_3}.  Sanders discussed tentatively the possibility of a $\chi^{(0)}<0$ state in media with inverted populations of energy levels produced by means similar to those used in maser and laser applications \cite{sanders_1988}, and Chiao \textit{et al}. predicted unequivocally a $\chi^{(0)}<0$ state in such systems \cite{chiao_1993,chiao_1994,chiao_1995_1,chiao_1995_2,chiao_1995_3}.  Apart from the unusual property of reducing, rather than increasing, the static mutual capacitance of two conductors if such a material were placed between them, the possibility of negative $\chi^{(0)}$---a qualitatively new parameter range for this important material property---opens up the possibility of new physical effects and technological capabilities.

For example, Sanders \cite{sanders_1988} and Chiao \textit{et al}. \cite{chiao_1994,chiao_1995_1,chiao_1995_2,chiao_1995_3} discussed theoretically how negative static electric susceptibility opens up the possibility of stable electrostatic levitation, which would be analogous in many respects to the magnetostatic levitation that is seen using diamagnetic or superconducting materials \cite{thomson_1847,braunbek_1939_1,braunbek_1939_2,arkadiev_1947}.  In particular, Chiao \textit{et al}. proposed explicitly that it should be possible to construct purely electrostatic charged particle traps, which would be very different in principle from Paul and Penning traps \cite{chiao_1994,chiao_1995_1,chiao_1995_2}.  However, Chiao \textit{et al}. also went on to predict theoretically that for a specific, typical case of the types of systems they considered (ammonia gas at a temperature of 180 K and a pressure of 1 Torr with population inversion maintained by a carbon dioxide pump laser), the value of $\chi^{(0)}$ is expected to be $\approx-3\times 10^{-4}$ \cite{chiao_1995_1,chiao_1995_2}; based on this relatively-small magnitude they concluded that condensed media with much larger magnitudes of the negative static electric susceptibility would need to be developed before such levitation effects become readily observable \cite{chiao_1995_1,chiao_1995_2}.  This is perhaps one reason why no attempts to observe experimentally the theoretically-predicted $\chi^{(0)}<0$ state, or the associated levitation phenomenon, have been reported using the systems considered by Sanders and Chiao \textit{et al}.

More recently, it was proposed that negative static electric susceptibility may be achieved also in completely different systems: active metamaterials \cite{castles_2020}.  Active metamaterials \cite{auzanneau_1999,tretyakov_2001,cummer_2015} utilize an internal source of power and, like the systems considered by Sanders and Chiao \textit{et al}., are not subject to the equilibrium thermodynamical argument of Landau \textit{et al}. concerning the restriction on the sign of $\chi^{(0)}$. In Ref.~\cite{castles_2020}, a design concept for active metamaterials with negative static electric susceptibility was proposed and preliminary experimental evidence in support of the general concept was reported.  Unlike the systems considered by Sanders and Chiao \textit{et al}., such metamaterials are readily realized at room temperature and pressure.  Further, they constitute a form of condensed matter for which it appears very likely that negative static electric susceptibility values with magnitudes much greater than the systems of Sanders and of Chiao \textit{et al}. are possible \cite{castles_2020}.

Given that: (1) the likely existence of negative static electric susceptibility in non-equilibrium materials raises the theoretical possibility of novel technological capabilities \cite{sanders_1986,sanders_1988,chiao_1994,chiao_1995_1,chiao_1995_2,chiao_1995_3}, (2) the magnitude of the negative static electric susceptibility that is achievable in non-equilibrium materials is expected to be crucial to the practical realization of such technologies \cite{chiao_1995_1,chiao_1995_2}, and (3) the practical development of condensed (meta)materials with negative static electric susceptibilities of relative large magnitudes is now well under way \cite{castles_2020}, it appears timely to seek to determine rigorously just how negative can one expect the negative static electric susceptibility to be made in non-equilibrium condensed materials. 

If Ref.~\cite{castles_2020} it was claimed---without full theoretical or experimental justification---that negative static electric susceptibility is possible throughout the range $-1<\chi^{(0)}<0$; herein we present the theoretical basis on which this claim was made.

\subsection{Scope} \label{sec:scope}

We consider the basic, conventional interpretations of $\chi^{(0)}$ and $\varepsilon^{(0)}$; that is, we consider the linear static electric susceptibility (polarization proportional to internal electric field) and linear static relative permittivity (electric displacement proportional to internal electric field) as they pertain to a nonrelativistic, macroscopic, and homogeneous sample of material under the action of a static electric field created by external test electrodes.  This interpretation corresponds, for example, to the original experiments of Cavendish \cite{cavendish_1921} and Faraday \cite{faraday_ERE_11_series_phil_trans}, to the meaning ascribed to the static electric susceptibility and permittivity in standard textbook accounts of the electrostatics of dielectrics (e.g., Chap. II of Ref.~\cite{ll8}), and to the definition of the permittivity according to recent ASTM standards \cite{astm_d150_18_2022}.

It should be noted that there are a number of scenarios where negative static electric susceptibility has been discussed or implied in the literature in relation to phenomena that do not correspond to this conventional interpretation.  For example, Kirzhnits \textit{et al}. have shown that static permittivity may be negative in the sense that, if spatial dispersion is taken into account, the longitudinal permittivity at zero frequency but nonzero wave vector % 'wave vector', not 'wavevector' according to AIP Style Manual
may exhibit negative values \cite{kirzhnits_1976,dolgov_1981,kirzhnits_1987,kirzhnitz_1989}.  However, this concerns the scenario where external sources of charge are located within the material itself, and, for the case of static fields and external test electrodes, i.e., the conventional case which corresponds to zero frequency and zero wave vector, Kirzhnits \textit{et al}. reaffirm the conclusion of Landau \textit{et al}. that the static permittivity of a material must exceed unity and hence the static electric susceptibility must be positive (if the body is in thermodynamic equilibrium).

We emphasize that we are interested in the static \textit{electric} susceptibility and permittivity; it is well known that the static \textit{magnetic} susceptibility may be either positive or negative, regardless of whether the body is in thermodynamic equilibrium or not (see, e.g., Ref.~\cite{ll8}, p.~106).  We also emphasize that we are interested in the \textit{static} electric susceptibility and permittivity; it is well known that the sign of the real part of the complex permittivity is subject to no theoretical restriction for nonzero frequencies, again, regardless of whether the body is in thermodynamic equilibrium or not (see, e.g., Ref.~\cite{ll8}, p.~274).

Finally, we note that we consider herein only isotropic media such that the static electric susceptibility---which is, in general, a real symmetric second-rank tensor---reduces to a real scalar, which we denote as $\chi^{(0)}$.  Thus, `negative static electric susceptibility' means, straightforwardly, that $\chi^{(0)}<0$ \footnote{Although we do not make use of the fact in what follows, a `negative static electric susceptibility' may be considered readily in the general, anisotropic, case to mean that one or more of the three principal components of the static electric susceptibility tensor are less than zero.}.    We use superscript `$(0)$' to denote explicitly a \textit{static} quantity.

To study the static electric susceptibility, as interpreted conventionally in this way, we use a simple model of insulators that is standard in the normal case of $\chi^{(0)}>0$.  To analyze the model we also use methods employed previously to a substantial extent in the case $\chi^{(0)}>0$.  The main novelty of our work is that new results are obtained by applying a generalized version of the method to the unusual parameter range $\chi^{(0)}<0$.

\subsection{A known lower bound on the value of \texorpdfstring{$\chi^{(0)}$}{} for non-equilibrium materials}

As noted above, the theoretical argument of Landau \textit{et al}. (\S14 of Ref.~\cite{ll8}) puts a lower bound of zero on the static electric susceptibility of materials that are in thermodynamic equilibrium.  A simple theoretical argument that puts a less stringent lower bound of $-1$ on the static electric susceptibility---but which applies manifestly to all materials, whether equilibrium or non-equilibrium---is provided by circuit theory, as follows \footnote{We are unaware of the origins of this type of argument to establish minimum bounds on $\varepsilon^{(0)}$ and $\chi^{(0)}$, but such an approach was brought to our attention by Ref.~\cite{tretyakov_2007}.}.

Consider an empty parallel plate capacitor of capacitance $C_0\,(>0)$ for which the plates are initially disconnected and a charge of $\pm q_0$ is deposited on the plates.  Now fill the volume between the capacitor plates with a homogeneous and isotropic insulator with static relative permittivity $\varepsilon^{(0)}$.  Suppose that the geometry of the capacitor is such that fringing fields are sufficiently small that they may be reasonably ignored.  By definition of $\varepsilon^{(0)}$ \cite{astm_d150_18_2022}, the capacitance of the filled capacitor is $C=\varepsilon^{(0)}C_0$.  Now let the capacitor plates be connected at time $t=0$ across a resistor of resistance $R\,(>0)$.  Suppose that the value of $R$ is such that the time constant $t_\text{c}=|RC|$ is large with respect to the time taken for polarization to be established in the material, and the resulting `quasi-static' behavior of the capacitor is dictated by $\varepsilon^{(0)}$.  %\textcolor{red}{[Think through this, make sure that it makes sense.]}
  Elementary circuit theory tells us that the charge on the capacitor as a function of time is $q(t)=q_0\exp\left[-t/\left(RC\right)\right]$.  For the normal case, $\varepsilon^{(0)}>1$ and therefore $C>0$, hence $q(t)$ decays exponentially, i.e., the capacitor discharges through the resistor.  However, for the hypothetical case $\varepsilon^{(0)}<0$, $C$ would be negative and an exponential, \textit{unbounded increase} of $q(t)$ would be predicted.  Since this indicates a system that is unstable, values $\varepsilon^{(0)}<0$, i.e., values $\chi^{(0)}<-1$, must be unphysical.  We note that the argument does not rely, at any point, on the material being in thermodynamic equilibrium, so it should apply to all materials whether or not they are in thermodynamic equilibrium \footnote{An alternative, briefer argument is to observe that the energy of a capacitor is given by $Q^2/\left(2C\right)$, which already indicates an instability for $C<0$, and hence for $\varepsilon^{(0)}<0$ and $\chi^{(0)}<-1$.  However, we believe that such energy-based arguments should, in general, be applied with some care to non-equilibrium materials that may utilize a source of power to maintain an electrostatic state that is otherwise energetically unfavorable.}.

This circuit theory result is of relatively little interest in relation to equilibrium materials because, in this case, the argument of Landau \textit{et al}. already provides a more stringent lower bound.  It is of much greater significance in relation to non-equilibrium materials because, in this case, it provides a lower bound where the argument of Landau \textit{et al}. does not apply.  Importantly, the circuit theory argument rules out most negative values of $\chi^{(0)}$ (viz., all values $<-1$) in non-equilibrium materials, and it already tells us that one cannot hope to obtain negative values with the large magnitudes readily seen in the positive case for equilibrium materials, even if condensed media are employed (many typical solids have values $\chi^{(0)}\sim 10^{0}$--$10^{3}$ \cite{kaye_laby_1973} and values of $\chi^{(0)}\sim 10^4$--$10^{5}$ may be observed just above the Curie point in materials that exhibit a ferroelectric phase \cite{hippel_1946,hippel_1954_2}).  On the other hand, it is essential to bear in mind that the circuit theory argument does not imply that any values in the interval $-1<\chi^{(0)}<0$ are necessarily possible for non-equilibrium materials. (Indeed, it is known, from Landau \textit{et al}.'s result, that this would be an incorrect inference in the case of equilibrium materials, to which the circuit theory argument applies equally).

To determine theoretically whether values in the interval $-1<\chi^{(0)}<0$ are actually possible for non-equilibrium materials, i.e., to determine the lower \textit{permissible} bound of $\chi^{(0)}$, it appears necessary to investigate a specific microscopic model of insulators.

\subsection{Model} \label{sec:model}

As presently conceived, materials with negative static electric susceptibility are composed of discrete entities with negative static polarizability.  To begin to model condensed media consisting of such entities---be they atoms, molecules, or meta-atoms---we assume herein a set of identical, point-like, dipolarizable entities that are located at fixed positions and respond linearly and isotropically to the local electric field: in static equilibrium, the dipole moment $\mathbf{p}^{(j)}$ ($\in\mathbb{R}^3$) of the $j$\textsuperscript{th} entity is given by $\mathbf{p}^{(j)}=\alpha\mathbf{E}^{(j)}$, where $\alpha$ ($\in\mathbb{R}$) is the static polarizability of each entity and $\mathbf{E}^{(j)}$ ($\in\mathbb{R}^3$) is the (`microscopic') electric field at the point $\mathbf{r}^{(j)}$ due to all sources except the $j$\textsuperscript{th} entity itself.  For a stable macroscopic medium composed of a cubic arrangement of such entities, the static electric susceptibility is also isotropic and described by scalar $\chi^{(0)}$ ($\in\mathbb{R}$) \footnote{It is, of course, also possible to obtain an isotropic electric susceptibility from entities with anisotropic static polarizability: for example, if the principal axes of the entities are randomly oriented.  However, for simplicity, we do not consider this case herein.}.

This model is, of course, precisely that which has been applied standardly, assuming $\alpha>0$, to the study of normal dielectrics with $\chi^{(0)}>0$, in which context it may be traced back to the work of Mossotti \cite{mossotti_1836,mossotti_1850}, Faraday \cite{faraday_ERE_14_series_phil_trans}, %I am referring here to paragraph 1679 of this reference, where Faraday discusses his `shot' model of dielectrics
 and Clausius \cite{clausius_1879} (though these authors envisioned small conducting inclusions of finite, nonzero, volume rather than dipolarizable entities that were assumed point-like from the outset).  The treatment of the usual case---i.e., the case $\alpha>0$ and $\chi^{(0)}>0$---that we believe is most relevant to our work is that of Allen \cite{allen_2004,allen_2004_inbook}.  Our task is to investigate the behavior of the model when $\alpha<0$.

\subsection{The Drude formula and the Clausius-Mossotti equation for the case of negative static polarizability}

If materials composed of entities with $\alpha<0$ are stable against self-polarization, then there is no reason to suppose that, in the weakly-interacting limit, the usual, basic expression $\chi^{(0)}=N\alpha/\varepsilon_0$ would not be as valid for the case $\alpha<0$ as it is for the case $\alpha>0$.  Here, $N$ ($\geqslant 0$) is the number of entities per unit volume and $\varepsilon_0$ is the permittivity of free space.  This expression is sometimes referred to as the Drude formula for susceptibility (see, e.g., Ref.~\cite{anderson_1951}) and it was assumed to be valid in the case of $\alpha<0$ by Chiao \textit{et al}. \cite{chiao_1993,chiao_1994,chiao_1995_1,chiao_1995_2}; in particular, Chiao \textit{et al}. used it to arrive at their theoretical prediction of the value of $\chi^{(0)}\approx-3\times 10^{-4}$ for pumped ammonia gas in Refs.~\cite{chiao_1995_1,chiao_1995_2}.  %The assumption, or approximation, that the entities do not interact mutually is well justified for the dilute atomic and molecular systems considered by Chiao \textit{et al}.

The Drude formula reflects the obvious fact that, if one wishes to achieve a negative $\chi^{(0)}$ of larger magnitude, it is natural to seek to employ entities with negative $\alpha$ of larger magnitude and/or arrangements of entities that are more densely packed, i.e., larger $N$ (hence the desire for condensed media).  Based on the work presented in Ref.~\cite{castles_2020}, it is clear that $\alpha$ can be negative and the combination $N\alpha$ can be of large magnitude in metamaterial systems.  However, as $|N\alpha|$ increases, the approximation that the entities may be treated as non-interacting---the assumption on which the derivation of the Drude formula is based---becomes increasingly inaccurate and mutual interactions between the entities must be taken into account.  Further, it must be borne in mind that, by applying the Drude formula to the $\alpha<0$ case at all, one is \textit{assuming} that the $\chi^{(0)}<0$ state exists; that is, one is assuming that the mutual interactions between the entities do not, in fact, make the system unstable for all values $\alpha<0$, no matter how small the value of $|N\alpha|$, in the macroscopic limit.  To our knowledge, there is no definitive way to justify this assumption theoretically without taking the mutual interactions into account explicitly and confirming whether the lower limit of stability is negative definite.

\begin{figure} \centering
\includegraphics[width=129mm]{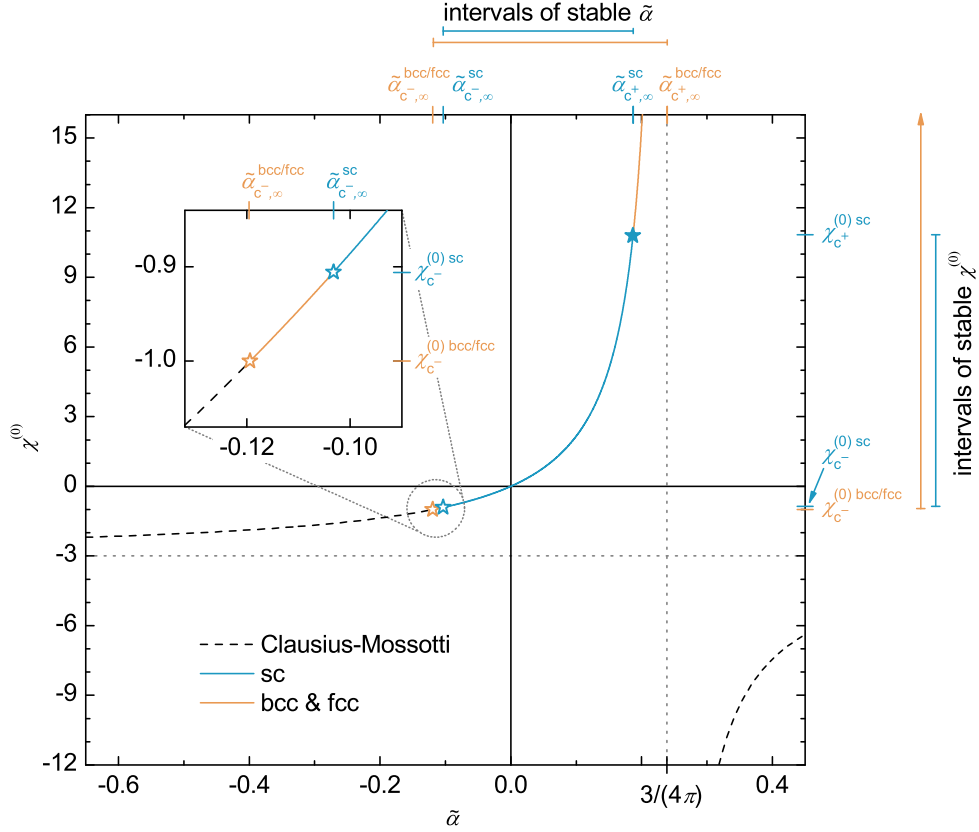}
\caption{The static electric susceptibility $\chi^{(0)}$ of the material as a function of the reduced static polarizability $\tilde\alpha$ of the dipolarizable entities from which the material is composed, according to the model.  In each of the sc, bcc, and fcc cases, the Clausius-Mossotti curve is followed, but only within a certain interval, $\tilde\alpha_{\text{c}^-,\,\infty}<\tilde\alpha<\tilde\alpha_{\text{c}^+,\,\infty}$.  For values of $\tilde\alpha$ such that $\tilde\alpha<\tilde\alpha_{\text{c}^-,\,\infty}$ or $\tilde\alpha>\tilde\alpha_{\text{c}^+,\,\infty}$ the system is unstable.  The lower permissible bounds for bcc and fcc are identical to within the accuracy of the numerical method (orange open star), but the lower permissible bound for sc is distinct  (blue open star).  It is seen that the intervals of stability extend into the region of negative parameters and hence negative values of $\chi^{(0)}$ are possible, according to the model, in each case.  The permissible lower bounds are not related to the horizontal asymptote of the Clausius-Mossotti curve (horizontal gray dotted line).} \label{fig:cm}
\end{figure}

Moving beyond the Drude formula, one arrives at the Clausius-Mossotti equation.  The Clausius-Mossotti equation accounts for the fact that the field at a given dipolarizable entity is not, in general, just the external field, but is the sum of the external field and the field due to the polarization of all the other entities in the material.  The assumptions of our model (discrete entities that are point-like, linearly-dipolarizable, at fixed positions, etc.), are exactly those often used in modern accounts of the derivation of the Clausius-Mossotti equation (see, e.g., Refs.~\cite{kittel_book,aspnes_1982}).  The equation states that, for media with certain symmetries (including cubic crystals with one entity per primitive cell, with which we are concerned herein) \cite{clausius_1879,hippel_1954_2,kittel_book}, the static electric susceptibility is related to the static polarizability as
\begin{equation} \label{eq:cm_reduced}
\chi^{(0)}=\frac{4\pi\tilde\alpha}{1-\frac{4\pi}{3}\tilde\alpha}.
\end{equation}
Here, we have chosen to write the equation in terms of the dimensionless variable $\tilde\alpha=N\alpha/\left(4\pi\varepsilon_0\right)$, which will be particularly convenient later.  A plot of $\chi^{(0)}$ as a function of $\tilde\alpha$ according to Eq.~(\ref*{eq:cm_reduced}) is included in Fig.~\ref*{fig:cm}.  As $\tilde\alpha\rightarrow \frac{3}{4\pi}$ from below, the Clausius-Mossotti equation predicts that $\chi^{(0)}\rightarrow\infty$.  This is sometimes referred to as the ``Lorentz $\frac{4\pi}{3}$ catastrophe'' and is associated with an instability towards self-polarization that has, for a long time, been considered a qualitative, if not precisely quantitative, model of the paraelectric to ferroelectric transition (see, e.g., Ref.~\cite{anderson_1951}).  %'paraelectric' not used by Anderson, but is used in IUPAC entry on "ferroelectric (antiferroelectric) transition" and source article for that entry: "Definitions of terms relating to phase transitions of the solid state (IUPAC Recommendations 1994)", Clarke, J.B.;Hastie, J.W.;Kihlborg, L.H.E.;Metselaar, R.;Thackeray, M.M., Pure and Applied Chemistry 1994, 66(3), 577
For $\tilde\alpha> \frac{3}{4\pi}$, the Clausius-Mossotti equation formally predicts negative values of $\chi^{(0)}$; however, this does not constitute a genuine prediction of a negative static electric susceptibility state.  Rather, the equation ceases to be valid in this case and a more sophisticated (nonlinear) model is required to describe the self-polarized state.  Accordingly, the conventional interval in which the Clausius-Mossotti equation is typically considered is $0\leq \tilde\alpha < \frac{3}{4\pi}$.

Since the Clausius-Mossotti equation is a useful description of dielectrics in the standard case of $0\leq \tilde\alpha < \frac{3}{4\pi}$, it is natural for us to ask whether, and to what extent, it remains valid in the case of $\tilde\alpha<0$.  It was noted as long ago as 1873 by Maxwell \cite{maxwell_1873_p_65} that there is no step in the derivation of the Clausius-Mossotti equation that requires the static polarizability to be positive for the derivation to be valid.  Therefore, if our model consists of entities that may assume negative values of $\tilde\alpha$ (as per the inverted media of Sanders \cite{sanders_1986,sanders_1988} and Chiao \textit{et al}. \cite{chiao_1994,chiao_1995_1,chiao_1995_2,chiao_1995_3}, or the metamaterials of Ref.~\cite{castles_2020}), one would expect, naively, that the Clausius-Mossotti equation will still provide the appropriate expression for $\chi^{(0)}$ for the model, as per Eq.~(\ref*{eq:cm_reduced}).  However, as we will now see, this is only partly true.

\subsubsection{The limitation of the Clausius-Mossotti equation for the case of negative static polarizability}

Since, unlike the case of $\tilde\alpha>0$, there is no divergence of the function $\chi^{(0)}$ for any $\tilde\alpha<0$ according to Eq.~(\ref*{eq:cm_reduced}), one would expect, naively, that Eq.~(\ref*{eq:cm_reduced}) should be applicable for all $\tilde\alpha<0$.  If this were the case, the minimum value of $\chi^{(0)}$ would be $\lim_{\tilde\alpha\rightarrow -\infty}\chi^{(0)}(\tilde\alpha)=-3$ (the horizontal asymptote of the Clausius-Mossotti curve as it is presented in Fig.~\ref*{fig:cm}).  However, this is clearly at odds with the circuit theory argument, discussed previously, which has shown that any value $\chi^{(0)}<-1$ is unstable.

To reconcile the fact that the naive result of the Clausius-Mossotti equation for $\tilde\alpha<0$ is at odds with the circuit theory argument, one may look at the case of a sc crystal \footnote{When we talk herein of sc, bcc, or fcc crystals, we mean, in all cases, crystals with one entity per primitive cell.} of $\tilde\alpha>0$ entities to remind us of a known limitation of the Clausius-Mossotti equation that exists even within the usual interval in which it is considered to be valid, i.e., within the interval $0\leq \tilde\alpha < \frac{3}{4\pi}$.  The derivation of the equation would appear to apply equally to sc, bcc, and fcc crystals, and, as discussed above, Eq.~(\ref*{eq:cm_reduced}) predicts that $\chi^{(0)}\rightarrow\infty$ as $\tilde\alpha\rightarrow \frac{3}{4\pi}$ from below in all three cases.  However, although it is not discussed explicitly in typical textbook accounts of the Clausius-Mossotti equation, it is, nevertheless, well established that a more-sophisticated analysis of the same model shows that, upon increasing $\tilde\alpha$ from zero, the sc case becomes unstable towards self polarization at a critical value $\tilde\alpha=\tilde\alpha_{\text{c}^+}^{\,\text{sc}}$ which is less than $\frac{3}{4\pi}$ \footnote{This has been articulated clearly by Allen in Refs.~\cite{allen_2004,allen_2004_inbook} and is related closely to the fact that, for a model consisting of entities with \textit{dipole moments of fixed magnitude that are free to rotate}, an antiferroelectric configuration, as opposed to a ferroelectric configuration, can be energetically preferred in the sc case \cite{sauer_1940,luttinger_1946}.} (the numerical value of $\tilde\alpha_{\text{c}^+}^{\,\text{sc}}$ will be considered in detail below).  What happens, more precisely, in the sc case is that, upon increasing $\tilde\alpha$ from zero, Eq.~(\ref*{eq:cm_reduced}) determines correctly the associated value of $\chi^{(0)}$, but only until the critical value $\tilde\alpha=\tilde\alpha_{\text{c}^+}^{\,\text{sc}}<\frac{3}{4\pi}$ is reached: for any value of $\tilde\alpha$ that is greater than $\tilde\alpha_{\text{c}^+}^{\,\text{sc}}$ the system is already unstable and the Clausius-Mossotti curve for the sc case is truncated (as shown in Fig.~\ref*{fig:cm}).  Accordingly, in the sc case for $\tilde\alpha>0$, the value of $\chi^{(0)}$ is not unbounded above, but has a finite upper bound determined by evaluating $\chi^{(0)}$ according to Eq.~(\ref*{eq:cm_reduced}) at $\tilde\alpha=\tilde\alpha_{\text{c}^+}^{\,\text{sc}}$.  Thus, there is clear precedent for the limited applicability of the Clausius-Mossotti equation within the normally-considered interval $0\leq\tilde\alpha<\frac{3}{4\pi}$, and for a finite \textit{upper} bound on $\chi^{(0)}$, for certain crystal structures.

As is generally assumed, and we will end up essentially reaffirming in detail below, there is no such truncation of the Clausius-Mossotti curve for the bcc and fcc cases with $\tilde\alpha>0$; within the model, the upper critical values for the static polarizability are $\tilde\alpha_{\text{c}^+}^{\,\text{bcc}}=\tilde\alpha_{\text{c}^+}^{\,\text{fcc}}=\frac{3}{4\pi}$ and, therefore, $\chi^{(0)}$ does indeed diverge to infinity as $\tilde\alpha\rightarrow \frac{3}{4\pi}$ from below.

For convenience, we may refer to an instability that occurs with a divergence of $\chi^{(0)}$---such as that which occurs for the bcc and fcc cases on increasing $\tilde\alpha$ from zero---as a `Type-I' instability, and an instability that occurs without a divergence of $\chi^{(0)}$---such as that which occurs for the sc case on increasing $\tilde\alpha$ from zero---as a `Type-II' instability.  Accordingly, to reconcile the fact that the minimum value of $\chi^{(0)}=-3$ that is naively predicted by the Clausius-Mossotti equation is at odds with the minimum value of $\chi^{(0)}=-1$ predicted by the circuit argument, we may hypothesize that, upon decreasing $\tilde\alpha$ from zero, there is always a Type-II instability that truncates the Clausius-Mossotti curve at or before $\chi^{(0)}\left(\tilde\alpha=-\frac{3}{8\pi}\right)=-1$.  Sections \ref*{sec:method} and \ref*{sec:results} of this paper amount, essentially, to a rigorous demonstration that this is indeed what happens, and to an accurate calculation of the values of $\tilde\alpha$ and $\chi^{(0)}$ at which the truncations occur, for the sc, bcc, and fcc cases.  Our primary methodology is a generalization of one of the methods used by Allen---which was applied in Refs.~\cite{allen_2004,allen_2004_inbook} to the sc case with $\alpha>0$---to include also the case of $\alpha<0$.

\subsection{The physical origin of the lower bound}

We believe that, in broad terms, the physical mechanism underlying the instability is the same for the $\tilde\alpha<0$ case as for the $\tilde\alpha>0$ case (which is essentially the same for Type I and Type II instabilities). That is, the physical mechanism of the instability in the $\tilde\alpha<0$ case is essentially the same as the well-known Lorentz $\frac{4\pi}{3}$ catastrophe model of the paraelectric to ferroelectric transition, which may be interpreted in the following way.  It is clear that, in the absence of external sources of electric fields, the state of the system where all of the entities are unpolarized, i.e., $\mathbf{p}^{(j)}=\mathbf{0}$ for all $j$, is a state of static equilibrium of the system; for, if the polarization of each entity is exactly zero, then, in the absence of external sources, there are no electric fields that may induce polarization.  The question is whether this state is a stable or unstable static equilibrium.  Loosely speaking, if the entities are arranged sufficiently close together or $|\alpha|$ is sufficiently large, then the system may be shown to be unstable;  in this case, any transient, non-zero, polarization of any of the entities would cause polarization of the other entities, which would, in turn, lead to further polarization which is, on average, of larger magnitude, etc., leading to a ``runaway condition" where polarization increases, in theory, without bound.  (Using a more-sophisticated, nonlinear model the self-polarization can be bounded and describe a ferroelectric or antiferroelectric state.)  Conversely, if the entities are arranged sufficiently far apart or $|\alpha|$ is sufficiently small, then the system may be shown to be stable; in this case, any transient, non-zero polarization of any of the entities would lead to only finite oscillations of the $\mathbf{p}^{(j)}$, and these would die away to zero if any damping whatsoever were present in the system \footnote{This type of instability is already exhibited by a system containing only two entities.  In this case, if the value of $\alpha$ in relation to the distance between the entities $d$ is such that $\alpha>2\pi\varepsilon_0 d^3$ then the system is unstable towards a self-polarized state.  For a pedagogical account of the two-entity system, and others, in the case $\alpha>0$, see Ref.~\cite{allen_2004_inbook}.  The two-entity system with $\alpha>0$ also appears, for example, as Problem 8 in Chapter 16 of Ref.~\cite{kittel_book}, where it is referred to as the ``ferroelectric criterion for atoms''.}.

\section{Method} \label{sec:method}

We will find that any set of two or more entities (assuming the entities are located at distinct positions) has an upper critical value of $\tilde\alpha$, which we denote $\tilde\alpha_{\text{c}^+}$, and a lower critical value of $\tilde\alpha$, which we denote $\tilde\alpha_{\text{c}^-}$, such that: if $\tilde\alpha_{\text{c}^-}<\tilde\alpha<\tilde\alpha_{\text{c}^+}$ then the system is stable, and if $\alpha<\tilde\alpha_{\text{c}^-}$ or $\alpha>\tilde\alpha_{\text{c}^+}$ then the system is unstable.  We seek to determine the values of $\tilde\alpha_{\text{c}^+}$ and $\tilde\alpha_{\text{c}^-}$ as they pertain to infinite sc, bcc, and fcc crystals, then use Eq.~(\ref*{eq:cm_reduced}) to determine the associated upper and lower bounds of $\chi^{(0)}$.

To determine the values of $\tilde\alpha_{\text{c}^+}$ and $\tilde\alpha_{\text{c}^-}$ as they pertain to infinite crystals, our primary methodology is to consider finite crystals of a systematic range of sizes and determine the infinite limit by extrapolation.  (We also summarize an alternative methodology in \S\ref*{sec:alternative_method}.)  This method has essentially been employed by Allen to investigate the sc case for $\tilde\alpha>0$ (see, in particular, Fig.~1 of Ref.~\cite{allen_2004}).  As we will see, if the method is suitably generalized to deal also with the case $\tilde\alpha<0$, and suitable care is taken to execute and analyze the extrapolation, it can lead to definitive results with well-defined accuracy for each of the sc, bcc, and fcc cases for both $\tilde\alpha>0$ and $\tilde\alpha<0$.

\subsection{General method for entities located at arbitrary positions}

We begin by establishing the method of determining $\tilde\alpha_{\text{c}^+}$ and $\tilde\alpha_{\text{c}^-}$ for a finite number of entities at arbitrary positions, and later consider positions that represent finite sc, bcc, and fcc crystals.  Consider a finite number $n_\text{tot}\geqslant 2$ of entities located at fixed position vectors $\mathbf{r}^{(j)}$ ($\in\mathbb{R}^3$), $j=1...n_\text{tot}$.  For now, the $\mathbf{r}^{(j)}$ may be considered to be entirely arbitrary, save that no two entities are coincidentally positioned, i.e., $\mathbf{r}^{(j)}\neq\mathbf{r}^{(l)}$ if $j\neq l$.  As per our model, in static equilibrium the dipole moment of the $j$\textsuperscript{th} entity, $\mathbf{p}^{(j)}$, is given by $\mathbf{p}^{(j)}=\alpha\mathbf{E}^{(j)}$, where $\alpha$ is the static polarizability of each entity and $\mathbf{E}^{(j)}$ is the electric field at the point $\mathbf{r}^{(j)}$ due to all sources except the $j$\textsuperscript{th} entity itself.  In the absence of additional sources of electric fields external to the $n_\text{tot}$ dipolarizable entities themselves (an external field being superfluous to the question of the intrinsic stability of the system of entities itself), we may write that, in static equilibrium,
\begin{equation} \label{eq:p_j}
\mathbf{p}^{(j)}=\alpha\sum_{l\neq j} \mathbf{E}^{(j,l)},
\end{equation}
where $\mathbf{E}^{(j,l)}$ ($\in\mathbb{R}^3$) is the field at $\mathbf{r}^{(j)}$ due to the polarization $\mathbf{p}^{(l)}$ of the $l\textsuperscript{th}$ entity and $\displaystyle\sum_{l\neq j}$ denotes the sum over all $l=1...n_\text{tot}$ except for the value $j$.  Using the standard expression for the electric field $\mathbf{E}$ at a displacement $\mathbf{r}$ from a point dipole $\mathbf{p}$, viz., $\mathbf{E} = \left[3\left(\mathbf{p}\cdot\mathbf{r}\right)\mathbf{r}-r^2\mathbf{p}\right]/\left(4\pi\varepsilon_0r^5\right)$, where $r=|\mathbf{r}|$, Eq.~(\ref*{eq:p_j}) becomes
\begin{equation} \label{eq:p_j_2}
\mathbf{p}^{(j)} = \alpha\sum_{l\neq j}\frac{3\left[\mathbf{p}^{(l)}\cdot\mathbf{R}^{(j,l)}\right]\mathbf{R}^{(j,l)}-\left[R^{(j,l)}\right]^2\mathbf{p}^{(l)}}{4\pi\varepsilon_0\left[R^{(j,l)}\right]^5}.
\end{equation}
Here, $\mathbf{R}^{(j,l)}=\mathbf{r}^{(j)}-\mathbf{r}^{(l)}$ is the displacement of the $j$\textsuperscript{th} entity with respect to the $l$\textsuperscript{th}, and $R^{(j,l)}=|\mathbf{R}^{(j,l)}|$.  Expressing distances in multiples of a unit distance $a>0$ (later, $a$ will be the length of the sc cell edge), such that $\mathbf{r}^{(j)}=a\mathbf{r}'^{(j)}$ and $\mathbf{R}^{(j,l)}=a\mathbf{R}'^{(j,l)}$, Eq.~(\ref*{eq:p_j_2}) may be rewritten 
\begin{equation} \label{eq:p_j_R}
\mathbf{p}^{(j)} = \tilde\alpha\sum_{l\neq j}\frac{3\left[\mathbf{p}^{(l)}\cdot\mathbf{R}'^{(j,l)}\right]\mathbf{R}'^{(j,l)}-\left[R'^{(j,l)}\right]^2\mathbf{p}^{(l)}}{\left[R'^{(j,l)}\right]^5},
\end{equation}
where $\tilde\alpha=\alpha/(4\pi\varepsilon_0 a^3)$ is the reduced, dimensionless, static polarizability. 

It is convenient to re-express the $n_\text{tot}$ vector equations of dimension three that are given by Eq.~(\ref*{eq:p_j_R}) as a single matrix equation of dimension $3n_\text{tot}$:
\begin{subequations} \label{eq:P}
\begin{align} 
\mathsf{P}&=\tilde\alpha\mathsf{M}\mathsf{P}, \label{eq:P_MP} \\
\text{or}\quad \left(\mathsf{I}-\tilde\alpha\mathsf{M}\right)\mathsf{P}	&= \mathsf 0. \label{eq:P_MP_2}
\end{align}
\end{subequations}
Here, $\mathsf{I}$ is the $3n_\text{tot}\times 3n_\text{tot}$ identity matrix, $\mathsf{0}$ is the $3n_\text{tot}\times 1$ zero column matrix, $\mathsf{P}$ is a $3n_\text{tot}\times 1$ column matrix
\[
\mathsf{P}=
\left[
\begin{array}{cccccccccc}
p^{(1)}_1 & p^{(1)}_2 & p^{(1)}_3 & p^{(2)}_1 & p^{(2)}_2 & p^{(2)}_3 & ... & p^{(n_\text{tot})}_1 & p^{(n_\text{tot})}_2 & p^{(n_\text{tot})}_3
\end{array}
\right]^\textrm{T}
\]
where $p^{(j)}_1$, $p^{(j)}_2$, and $p^{(j)}_3$ are the Cartesian components of $\mathbf{p}^{(j)}$, and $\mathsf{M}$ is a $3n_\text{tot}\times 3n_\text{tot}$ matrix that may be most-conveniently specified in terms of $3\times 3$ sub-matrices $\mathsf{M}^{(i,j)}$,
\begin{equation} \label{eq:M_matrix}
\mathsf{M}=
\begin{bmatrix}
\mathsf{0}_3 & \mathsf{M}^{(1,2)} & \mathsf{M}^{(1,3)} & \dots & \dots & \mathsf{M}^{(1,n_\text{tot})} \\
\mathsf{M}^{(2,1)} & \mathsf{0}_3 & \mathsf{M}^{(2,3)} & \dots & \dots &\mathsf{M}^{(2,n_\text{tot})}\\
\mathsf{M}^{(3,1)} & \mathsf{M}^{(3,2)} & \mathsf{0}_3 & \dots & & \vdots\\
\vdots & \vdots & \vdots & \ddots & \vdots & \vdots \\
\vdots & \vdots & & \dots & \mathsf{0}_3 & \mathsf{M}^{(n_\text{tot}-1,n_\text{tot})} \\
\mathsf{M}^{(n_\text{tot},1)} & \mathsf{M}^{(n_\text{tot},2)} & \dots & \dots & \mathsf{M}^{(n_\text{tot},n_\text{tot}-1)} & \mathsf{0}_3
\end{bmatrix},
\end{equation}
where $\mathsf{0}_3$ is the $3\times 3$ zero matrix and
\begin{equation} \label{eq:M_3_matrix}
\left[\mathsf{M}^{(j,l)}\right]_{\beta\gamma}=\frac{3\,R'^{(j,l)}_\beta R'^{(j,l)}_\gamma-\left[R'^{(j,l)}\right]^2\delta_{\beta\gamma}}{\left[R'^{(j,l)}\right]^5},\quad \text{for}\quad\beta,\gamma =1,2,3.
\end{equation}
Here, $\delta_{\beta\gamma}$ denotes Kronecker's delta, $R'^{(j,l)}_1$, $R'^{(j,l)}_2$, and $R'^{(j,l)}_3$ are the Cartesian components of $\mathbf{R}'^{(j,l)}$, and, in terms of the Cartesian components, $R'^{(j,l)}=\sqrt{\left[R'^{(j,l)}_1\right]^2+\left[R'^{(j,l)}_2\right]^2+\left[R'^{(j,l)}_3\right]^2}$.

It is clear that the state $\mathsf{P}=\mathsf{0}$, i.e., the state where all of the entities are unpolarized, is always a solution of Eq.~(\ref*{eq:P}) and hence always an equilibrium state of the system.  This solution may be stable or unstable depending on the values of $\tilde\alpha$ and $\mathsf{M}$, i.e., depending on the static polarizability and the relative positions of the entities.  The precise criterion for stability of the system is as follows: the system is stable if all the eigenvalues of matrix $\left(\mathsf{I}-\tilde\alpha\mathsf{M}\right)$ are positive [note that the eigenvalues are real because $\left(\mathsf{I}-\tilde\alpha\mathsf{M}\right)$ is a real-symmetric matrix], and the system is unstable if any of the eigenvalues of $\left(\mathsf{I}-\tilde\alpha\mathsf{M}\right)$ are negative. 

That this is the precise criterion for stability of the system may be derived using a number of different arguments. Perhaps the simplest is to note that, although we are considering entities that may exploit a source of power to maintain an unnatural state of polarization, we can, nevertheless, consider the equilibrium condition, Eq.~(\ref*{eq:P}), as arising formally from the minimization of an energy function $U:\mathbb{R}^{3n_\text{tot}}\rightarrow\mathbb{R}$
\begin{equation}
U(\mathsf{P})=\frac{1}{2\left|\tilde\alpha\right|}\mathsf{P}^\text{T}\left(\mathsf{I}-\tilde\alpha\mathsf{M}\right)\mathsf{P},
\end{equation}
whereupon Eq.~(\ref*{eq:P}) arises from the condition $\partial U/\partial \mathsf {P}=\mathsf{0}$.  The matrix $\left(\mathsf{I}-\tilde\alpha\mathsf{M}\right)$ is thus identified as the Hessian matrix of the system and, according to standard results in mathematical analysis (see, e.g., Ref.~\cite{korner_companion}), the point $\mathsf{P}=\mathsf 0$ is a minimum of $U$ if $\left(\mathsf{I}-\tilde\alpha\mathsf{M}\right)$ is positive definite, i.e., if all its eigenvalues are positive.  If $\mathsf{P}=\mathsf 0$ is a minimum of the energy then it is a stable configuration of the system.  Similarly, the point $\mathsf{P}=\mathsf 0$ is a saddle or a maximum of $U$ if any of the eigenvalues of $\left(\mathsf{I}-\tilde\alpha\mathsf{M}\right)$ are negative.  If $\mathsf{P}=\mathsf 0$ is a saddle or a maximum of the energy, then the system is unstable along one or more directions in $\mathsf{P}$-space.

Since $\mathsf{M}$ is a real, symmetric, traceless matrix (of dimension $\geqslant 6$ since we are considering $n_\text{tot}\geqslant 2$) which is not equal to the $3n_\text{tot}\times 3n_\text{tot}$ zero matrix, we know that it has at least one positive (definite) eigenvalue and at least one negative (definite) eigenvalue, and the condition for stability may be restated most simply as follows: the system is stable for values of $\tilde\alpha$ such that $\tilde\alpha_{\text{c}^-}<\tilde\alpha<\tilde\alpha_{\text{c}^+}$, where $\tilde\alpha_{\text{c}^-}=1/\lambda_\text{min}<0$ and $\tilde\alpha_{\text{c}^+}=1/\lambda_\text{max}>0$.  Here, $\lambda_\text{min}$ and $\lambda_\text{max}$ are the minimum and maximum eigenvalues of $\mathsf{M}$ respectively.  Conversely, if $\tilde\alpha<\tilde\alpha_{\text{c}^-}$ or $\tilde\alpha>\tilde\alpha_{\text{c}^+}$ then the system is unstable.

Thus, the question of stability of a given system becomes essentially a question of constructing the matrix $\mathsf{M}$ for that system (which depends only on the relative positions of the entities) and calculating its maximum and minimum eigenvalues.  This stability criterion is a generalization---to include also the $\alpha<0$ case---of that presented, and derived by a somewhat similar argument, for the $\alpha>0$ case in Refs.~\cite{allen_2004,allen_2004_inbook}.

\subsection{Application to arrays forming finite sc, bcc, and fcc crystals}

Having established the stability criterion for entities located at arbitrary positions, we now proceed to specify explicitly the locations of the entities for sc, bcc, and fcc finite crystals used in the study.  At this stage, it is convenient to switch from labeling the entities with a single index $j$ (with position vectors $\mathbf{r}^{(j)}$, etc.) to labeling them with an ordered triple of integers $(u_1,u_2,u_3)$.  In this notation, finite crystals may be created via the usual approach \cite{ashcroft_mermin_p_66} of locating entities at the sets of (reduced) position vectors $S_n=\left\{\tilde{\mathbf{r}}^{(u_1,u_2,u_3)}:u_1,u_2,u_3=0...n-1\right\}$, where $\tilde{\mathbf{r}}^{(u_1,u_2,u_3)}=u_1\tilde{\mathbf{a}}_1+u_2\tilde{\mathbf{a}}_2+u_3\tilde{\mathbf{a}}_3$.  Here, $\tilde{\mathbf{a}}_1$, $\tilde{\mathbf{a}}_2$, and $\tilde{\mathbf{a}}_3$ are the (reduced) primitive lattice vectors of the associated Bravais lattice.  For the sc case: $\tilde{\mathbf{a}}_1^\text{sc}=\hat{\mathbf{x}}_1$, $\tilde{\mathbf{a}}_2^\text{sc}=\hat{\mathbf{x}}_2$, and $\tilde{\mathbf{a}}_3^\text{sc}=\hat{\mathbf{x}}_3$, where $\hat{\mathbf{x}}_1$, $\hat{\mathbf{x}}_2$, and $\hat{\mathbf{x}}_3$ are the (dimensionless) unit vectors of a conventional Cartesian system. For the bcc case: $\tilde{\mathbf{a}}_1^\text{bcc}=(1/2^{2/3})\left(-\hat{\mathbf{x}}_1+\hat{\mathbf{x}}_2+\hat{\mathbf{x}}_3\right)$, $\tilde{\mathbf{a}}_2^\text{bcc}=(1/2^{2/3})\left(\hat{\mathbf{x}}_1-\hat{\mathbf{x}}_2+\hat{\mathbf{x}}_3\right)$, and $\tilde{\mathbf{a}}_3^\text{bcc}=(1/2^{2/3})\left(\hat{\mathbf{x}}_1+\hat{\mathbf{x}}_2-\hat{\mathbf{x}}_3\right)$.  For the fcc case: $\tilde{\mathbf{a}}_1^\text{fcc}=(1/2^{1/3})\left(\hat{\mathbf{x}}_2+\hat{\mathbf{x}}_3\right)$, $\tilde{\mathbf{a}}_2^\text{fcc}=(1/2^{1/3})\left(\hat{\mathbf{x}}_3+\hat{\mathbf{x}}_1\right)$, and $\tilde{\mathbf{a}}_3^\text{fcc}=(1/2^{1/3})\left(\hat{\mathbf{x}}_1+\hat{\mathbf{x}}_2\right)$.  Each set $S_n$ describes a finite crystal consisting of an $n\times n\times n$ array of entities (total number of entities $n_\text{tot}=n^3$) with the overall shape of a rhombohedron (more specifically, a cube in the sc case).

We have chosen normalization factors, $1/2^{2/3}$ and $1/2^{1/3}$ for the bcc and fcc cases respectively, such that the primitive cells (and not the conventional cells) are of unit volume.  This means that the number density is the same in each of the sc, bcc, and fcc cases and given by $N=1/a^3$, which is convenient for our purposes.  In particular, the expression for $\tilde\alpha$ used in this section, $\tilde\alpha=\alpha/(4\pi\varepsilon_0 a^3)$, is consistent with the expression $\tilde\alpha=N\alpha/(4\pi\varepsilon_0)$ used in Section~\ref*{sec:model} and Eq.~(\ref*{eq:cm_reduced}), and the resulting values of $\tilde\alpha_{\text{c}^\pm}$ for the sc, bcc, and fcc cases may be inserted into Eq.~(\ref*{eq:cm_reduced}) and compared in a like-for-like fashion.

We used Python to generate the $\mathsf{M}$-matrices, calculate $\lambda_\text{min}$ and $\lambda_\text{max}$, and hence determine the critical values of the static polarizability $\tilde\alpha_{\text{c}^+,\,n}$ and $\tilde\alpha_{\text{c}^-,\,n}$ for a given $n\times n\times n$ crystal described by a given set $S_n$.  The Python code is included in the Supplemental Material \cite{supp_aip}.  In each of the sc, bcc, and fcc cases, $\tilde\alpha_{\text{c}^+,\,n}$ and $\tilde\alpha_{\text{c}^-,\,n}$ were calculated for $n=2,3,...,n_\text{max}$, with $n_\text{max}=27$ (thus, the largest crystals contained $27^3=19,683$ dipolarizable entities).  The value $n_\text{max}=27$ was used as it was found to be the largest value for which our code could be run on the system with 64~GB RAM that we had readily available and, as shown below, it already provides results that are sufficiently conclusive for our purposes.

We may note that the method takes into account all interactions between every pair of entities in the crystal (and does not, for example, assume only nearest neighbor interactions).

To determine the critical values of the static polarizability in the macroscopic limit, i.e., in the limit of infinite sc, bcc, and fcc crystals we calculate the limiting values $\tilde\alpha_{\text{c}^+,\,\infty}=\lim_{n\rightarrow\infty}\tilde\alpha_{\text{c}^+,\,n}$ and $\tilde\alpha_{\text{c}^-,\,\infty}=\lim_{n\rightarrow\infty}\tilde\alpha_{\text{c}^-,\,n}$ by extrapolation, as detailed below.

\section{Results} \label{sec:results}

\subsection{Critical values of the reduced static polarizability and static electric susceptibility}

\begin{figure} \centering
\includegraphics[width=86mm]{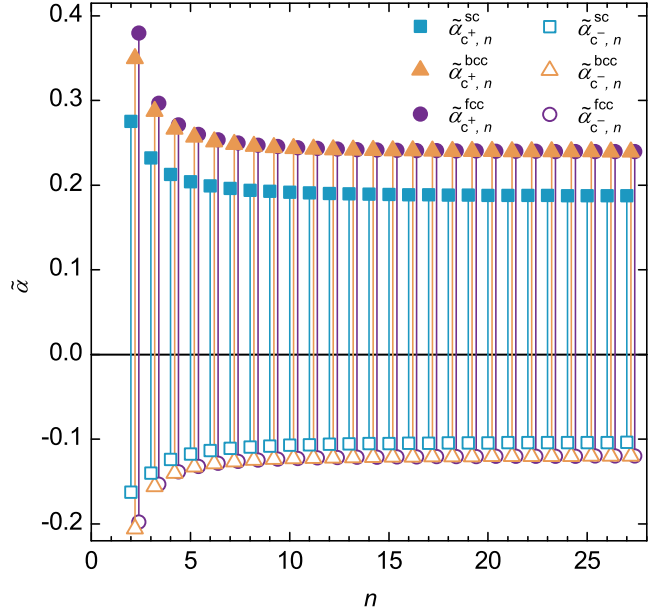}
\caption{The positive and negative critical values of the reduced static polarizability, $\tilde\alpha_{\text{c}^+,\,n}$ and $\tilde\alpha_{\text{c}^-,\,n}$ respectively, for sc, bcc, and fcc finite crystals as a function of $n$, for $n=2,3,4, ..., 27$ (total number of entities $n_\text{tot}=2^3, 3^3, 4^3, ..., 27^3$). The vertical lines denote the intervals of reduced static polarizability $\tilde\alpha_{\text{c}^-,\,n}<\tilde\alpha<\tilde\alpha_{\text{c}^+,\,n}$ that are stable in each case.  %The data for the bcc and fcc cases have been offset in the horizontal direction for clarity.
For clarity, the data sets for bcc and fcc have been displayed with artificial offsets in the horizontal direction of $n=0.2$ and $n=0.4$ respectively.
} \label{fig:basic}
\end{figure}

Raw data for the values of $\tilde\alpha_{\text{c}^-,\,n}$ and $\tilde\alpha_{\text{c}^+,\,n}$ in each of the sc, bcc, and fcc cases are given in Table~S1 of the Supplemental Material \cite{supp_aip}.  Plots of $\tilde\alpha_{\text{c}^-,\,n}$ and $\tilde\alpha_{\text{c}^+,\,n}$ for $n=2,3,...,n_\text{max}$ are shown in Fig.~\ref*{fig:basic}.
It is seen from the plots that: (1) In each of the sc, bcc, and fcc cases, the values of $\tilde\alpha_{\text{c}^+,\,n}$ decrease with increasing $n$ and the values of $\tilde\alpha_{\text{c}^-,\,n}$ increase with increasing $n$. (For the case of $\tilde\alpha^\text{\,sc}_{\text{c}^+,\,n}$, the decreasing behavior was essentially noted in \cite{allen_2004}, although there it was characterized as an increase in the associated eigenvalue, which amounts to the same thing, since $\tilde\alpha_{\text{c}^+}=1/\lambda_\text{max}$.)  Thus, the interval of stability of $\tilde\alpha$ is reduced in both its positive and negative extents as $n$ increases.  This makes intuitive physical sense; it would seem reasonable to expect that the addition of more dipolarizable entities can, loosely speaking, only serve to increase the overall amount of mutual interaction within the system, causing a greater tendency towards instability.  (2) In all cases, the sequences $\tilde\alpha_{\text{c}^\pm,\,n}$, $n=2,3,4,...$ appear to converge to finite, non-zero values. (We already know, by the very fact that the dielectric state exists for the case of positive static polarizability within this model, that the sequences $\tilde\alpha_{\text{c}^+,\,n}$, $n=2,3,4,...$ do not converge to zero for large $n$, and the qualitative behavior of the magnitude of sequences $\tilde\alpha_{\text{c}^-,\,n}$, $n=2,3,4,...$ appears to be very similar in this respect.) (3) In both the positive and negative cases, it appears that the values of $\tilde\alpha^\text{\,bcc}_{\text{c},\,n}$ and $\tilde\alpha^\text{\,fcc}_{\text{c},\,n}$ become increasingly coincident with each other as $n$ increases, whereas the values of $\tilde\alpha^\text{\,sc}_{\text{c},\,n}$ remain significantly different.

\begin{figure} \centering
\includegraphics[width=86mm]{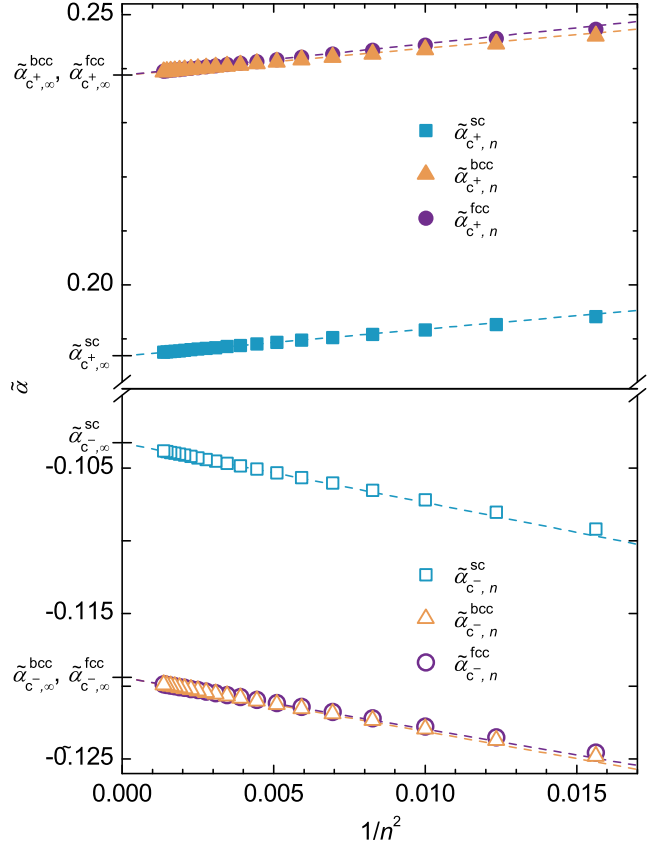}
\caption{Plots of the points $(1/n^2,\tilde\alpha_{\text{c}^+,\,n})$ are seen to be approximately linear (for clarity, only data for $n\geqslant 8$ are shown).  The extrapolated values in the limit $n\rightarrow\infty$ are obtained as the vertical-intercepts of straight lines through the two largest-$n$ data points ($n=26$ and $n=27$) in each case.
} \label{fig:extrap}
\end{figure}

These aspects of the data are illuminated further by plots of the points $(1/n^2,\tilde\alpha_{\text{c}^\pm,\,n})$, as shown in Fig.~\ref*{fig:extrap}.  The plots indicate also how the limiting values $\tilde\alpha_{\text{c}^\pm,\,\infty}$ may be determined by extrapolation.
It can be seen that the plots are approximately linear in each case; this enables the best estimates for the values of $\tilde\alpha_{\text{c}^\pm,\,\infty}=\lim_{n\rightarrow\infty}\tilde\alpha_{\text{c}^\pm,\,n}$ to be determined as the vertical intercepts of linear fit lines. Since the data becomes linear to an increasing level of accuracy as $n$ is increased, we use the two largest-$n$ data points available, $n=n_\text{max}-1=26$ and $n=n_\text{max}=27$, to construct the extrapolation line and determine what we believe to be our best estimates of the limiting values using this method.  Explicitly, the formula for the extrapolated values is thus
\begin{equation} \label{eq:extrap} %see LBXXX p.117
\tilde\alpha_{\text{c}^\pm,\,\infty}=\frac{n_\text{max}^2\tilde\alpha_{\text{c}^\pm,\,n_\text{max}}-\left(n_\text{max}-1\right)^2\tilde\alpha_{\text{c}^\pm,\,n_\text{max}-1}}{2n_\text{max}-1}.
\end{equation}
The values obtained by this method are listed in Table~\ref*{tab:data}, where they are quoted to three significant figures.
\begin{table}
\caption{\label{tab:data} Extrapolated critical values of the reduced static polarizability $\tilde\alpha_{\text{c}^\pm,\,\infty}$, and the associated critical values of the static electric susceptibility $\chi^{(0)}_{\text{c}^\pm}$, for sc, bcc, and fcc infinite crystals, stated to three significant figures.  Where available, the corresponding  exact values (expected or inferred) are indicated in square brackets.}
\begin{ruledtabular}
\begin{tabular}{c|ddd}
& \multicolumn{1}{c}{sc} & \multicolumn{1}{c}{bcc} & \multicolumn{1}{c}{fcc}\\
\hline
$\tilde\alpha_{\text{c}^+,\,\infty}\quad$	&0.187		& 0.239 \left[\frac{3}{4\pi}\right]		&	0.239 \left[\frac{3}{4\pi}\right] \\
$\tilde\alpha_{\text{c}^-,\,\infty}\quad$	&-0.103		& -0.119 \left[-\frac{3}{8\pi}\right]	&	-0.119 \left[-\frac{3}{8\pi}\right] \\
\hline
$\chi^{(0)}_{\text{c}^+}\quad						$	&10.8			& \multicolumn{1}{c}{\quad\quad$\left[\infty\right]$}		& \multicolumn{1}{c}{\quad\quad$\left[\infty\right]$} \\
$\chi^{(0)}_{\text{c}^-}\quad						$	&-0.906	&	-1.00 \left[-1\right]	&	-1.00 \left[-1\right]
\end{tabular}
\end{ruledtabular}
\end{table}

As discussed above, we expect the Clausius-Mossotti equation to remain valid for values of $\tilde\alpha$ that are within the stable interval $\tilde\alpha_{\text{c}^-}<\tilde\alpha<\tilde\alpha_{\text{c}^+}$.  Therefore, the critical values of the static electric susceptibility $\chi^{(0)}_{\text{c}^\pm}$ may be determined by inserting the values of $\tilde\alpha_{\text{c}^\pm,\,\infty}$ into Eq.~(\ref*{eq:cm_reduced}).  The values of $\chi^{(0)}_{\text{c}^\pm}$ thus obtained are listed also in Table~\ref*{tab:data} to three significant figures.  The results of the study are summarized in Fig.~\ref*{fig:cm}.

\subsection{Accuracy of the numerically-determined critical values} \label{sec:accuracy}

There are a number of ways by which we may consider the accuracy of the results determined via the above method. 

\subsubsection{Accuracy via comparison against assumed exact values for the cases of bcc and fcc with \texorpdfstring{$\tilde\alpha>0$}{}} \label{sec:acc_comp}

It appears reasonable to assume that the true model values (by which we mean the values predicted by the model, if we were able to solve it exactly as opposed to numerically) of $\tilde\alpha^\text{\,bcc}_{\text{c}^+,\,\infty}$ and $\tilde\alpha^\text{\,fcc}_{\text{c}^+,\,\infty}$ are exactly $\frac{3}{4\pi}$, as per the vertical asymptote in the Clausius-Mossotti expression, Eq.~(\ref*{eq:cm_reduced}); our numerical results for these values are $\tilde\alpha^\text{\,bcc}_{\text{c}^+,\,\infty}=0.238752$ and $\tilde\alpha^\text{\,fcc}_{\text{c}^+,\,\infty}=0.238756$ to six decimal places, which approximate $\frac{3}{4\pi}$ to an accuracy of 0.008\,\% and 0.010\,\% respectively.  Based on this, it appears reasonable to argue that, in general, the uncertainty in determining the values of $\tilde\alpha_{\text{c}^\pm,\,\infty}$ via our method is probably 0.01\,\% to the nearest order of magnitude (when using $n_\text{max}=27$); that is, we may assume an uncertainty of the order $0.01\,\%$ also on our values of $\tilde\alpha^\text{\,sc}_{\text{c}^+,\,\infty}$, $\tilde\alpha^\text{\,sc}_{\text{c}^-,\,\infty}$, $\tilde\alpha^\text{\,bcc}_{\text{c}^-,\,\infty}$, and $\tilde\alpha^\text{\,fcc}_{\text{c}^-,\,\infty}$.  Using the standard method for propagation of uncertainties, uncertainties of the order $0.01\,\%$ on $\tilde\alpha^\text{\,sc}_{\text{c}^+,\,\infty}$, $\tilde\alpha^\text{\,sc}_{\text{c}^-,\,\infty}$, $\tilde\alpha^\text{\,bcc}_{\text{c}^-,\,\infty}$, and $\tilde\alpha^\text{\,fcc}_{\text{c}^-,\,\infty}$, lead, via Eq.~(\ref*{eq:cm_reduced}), to uncertainties of the order of 0.01\,\% also on the values of $\chi^{(0)\,\text{sc}}_{\text{c}^+}$, $\chi^{(0)\,\text{sc}}_{\text{c}^-}$, $\chi^{(0)\,\text{bcc}}_{\text{c}^-}$, and $\chi^{(0)\,\text{fcc}}_{\text{c}^-}$ (more precisely, an uncertainty of 0.01\,\% on $\tilde\alpha^\text{\,sc}_{\text{c}^+,\,\infty}$ would lead to an uncertainty of 0.05\,\% on $\chi^{(0)\,\text{sc}}_{\text{c}^+}$, and uncertainties of 0.01\,\% on $\tilde\alpha^\text{\,sc}_{\text{c}^-,\,\infty}$, $\tilde\alpha^\text{\,bcc}_{\text{c}^-,\,\infty}$, and $\tilde\alpha^\text{\,fcc}_{\text{c}^-,\,\infty}$ would lead to uncertainties of 0.007\,\% on $\chi^{(0)\,\text{sc}}_{\text{c}^-}$, $\chi^{(0)\,\text{bcc}}_{\text{c}^-}$, and $\chi^{(0)\,\text{fcc}}_{\text{c}^-}$).  Therefore, we may conclude that the values listed in Table \ref*{tab:data} are almost certainly accurate as-quoted to three significant figures, and, if desired, values quoted to a larger number of significant figures with uncertainties of the order of 0.01\,\% may be readily generated from the raw data (provided in the Supplemental Material \cite{supp_aip}).  In particular: to a larger number of significant figures, we find $\chi^{(0)\,\text{bcc}}_{\text{c}^-}=-1.00009$ and $\chi^{(0)\,\text{fcc}}_{\text{c}^-}=-1.00008$.  Given the above-estimated uncertainty of the order of 0.01\,\%, these values are consistent with, and very suggestive of, the true model values for these quantities both being exactly $-1$.

Working backwards at this point to cross check, if we believe that the true model values of $\chi^{(0)\,\text{bcc}}_{\text{c}^-}$ and $\chi^{(0)\,\text{fcc}}_{\text{c}^-}$ are both $-1$, then the true model values of $\tilde\alpha^\text{\,bcc}_{\text{c}^-,\,\infty}$, and $\tilde\alpha^\text{\,fcc}_{\text{c}^-,\,\infty}$ must, from Eq.~(\ref*{eq:cm_reduced}), be exactly $-\frac{3}{8\pi}$.  Our numerical results for these values are $\tilde\alpha^\text{\,bcc}_{\text{c}^-,\,\infty}=-0.119382$ and $\tilde\alpha^\text{\,fcc}_{\text{c}^-,\,\infty}=-0.119381$ to six decimal places, which may be seen to approximate $-\frac{3}{8\pi}$ to an accuracy of 0.013\,\% and 0.012\,\% respectively.  This appears consistent with our previous assumption that the uncertainty on all the numerical values of $\tilde\alpha_{\text{c}^\pm,\,\infty}$ is of the order 0.01\,\%.

\subsubsection{Accuracy via inductive reasoning}

An alternative way to assess the accuracy of our numerical results is as follows.  Noting, from Fig.~\ref*{fig:basic} and the raw data tabulated in the Supplemental Material \cite{supp_aip}, that the sequences $\tilde\alpha_{\text{c}^+,\,n}$, $n=2,3,4,...,27$, decrease monotonically with increasing $n$ in each of the sc, bcc, and fcc cases, and assuming that this trend continues for all $n$ (i.e., applying `inductive reasoning'), then any particular value of $\tilde\alpha_{\text{c}^+,\,n}$ for a given crystal structure provides an upper bound on $\tilde\alpha_{\text{c}^+,\,\infty}$ for that crystal structure (note that we are referring here to an upper bound on the numerically-determined value of $\tilde\alpha_{\text{c}^+,\,\infty}$, which is itself an upper bound on the stable value of $\tilde\alpha$).  Accordingly, the most stringent upper bounds on the values of $\tilde\alpha_{\text{c}^+,\,\infty}$ that may be determined from the data in this way are given by $\tilde\alpha_{\text{c}^+,\,\infty}^\text{ub}=\tilde\alpha_{\text{c}^+,\,n_\text{max}}$ (with, in our case, $n_\text{max}=27$).

To determine lower bounds on the numerically-determined values of $\tilde\alpha_{\text{c}^+,\,\infty}$, we first recall that our extrapolation to determine the best-estimate values of $\tilde\alpha_{\text{c}^\pm,\,\infty}$, as per Fig.~\ref*{fig:extrap} and Eq.~(\ref*{eq:extrap}), used a straight line fit through the points $(1/n^2,\tilde\alpha_{\text{c}^+,\,n})$.  Here, the exponent of two in $1/n^2$ was chosen `by hand' to produce the most-linear plot.  If, instead of an exponent of two, we choose an exponent of one and, hence, plot instead the points $(1/n,\tilde\alpha_{\text{c}^+,\,n})$, we may observe that the graph is increasing and convex in each of the sc, bcc, and fcc cases (see Fig.~S1 of the Supplemental Material \cite{supp_aip}).  The choice of an exponent of one is somewhat arbitrary; the requirement is simply that a convex graph is produced.  Again, assuming this trend remains true, i.e., the plot remains increasing and convex, not just for $n=2...27$ but for all $n$, then the vertical intercept of a straight line through any two data points provides a lower bound on $\tilde\alpha_{\text{c}^+,\,\infty}$.  Accordingly, the most stringent lower bounds on the values of $\tilde\alpha_{\text{c}^+,\,\infty}$ that may be determined in this way from the data available are given by the vertical intercepts of the straight lines through the points $(1/n,\tilde\alpha_{\text{c}^+,\,n})$ with $n=n_\text{max}-1=26$ and $n=n_\text{max}=27$.  Explicitly, the formula for this procedure is thus
\begin{equation} \label{eq:extrap_plus_lower} %see LBXXX p.117
\tilde\alpha_{\text{c}^+,\,\infty}^\text{lb}=n_\text{max}\tilde\alpha_{\text{c}^+,\,n_\text{max}}-\left(n_\text{max}-1\right)\tilde\alpha_{\text{c}^+,\,n_\text{max}-1}.
\end{equation}

Similarly, with regard to $\tilde\alpha_{\text{c}^-,\,\infty}$, we may observe that the sequence $\tilde\alpha_{\text{c}^-,\,n}$, $n=2,3,4,...,27$ increases monotonically for each type of crystal structure and we may argue that lower bounds on the values of $\tilde\alpha_{\text{c}^-,\,\infty}$ are given by $\tilde\alpha_{\text{c}^-,\,\infty}^\text{lb}=\tilde\alpha_{\text{c}^-,\,n_\text{max}}$ in each case.  To determine upper bounds on the values of $\tilde\alpha_{\text{c}^-,\,\infty}$, we observe that plots of the points $(1/n,\tilde\alpha_{\text{c}^-,\,n})$ are decreasing and concave in each case, with the caveat that only points for which $n\geqslant 5$ are included in the sc case (see Fig.~S1 of the Supplemental Material \cite{supp_aip}).  Therefore, we may argue, analogously to above, that the most stringent upper bounds on the values of $\tilde\alpha_{\text{c}^-,\,\infty}$ that may be determined by this method from the data available are given by the vertical intercepts of the straight lines through the points $(1/n,\tilde\alpha_{\text{c}^-,\,n})$ with $n=n_\text{max}-1=26$ and $n=n_\text{max}=27$.  Explicitly, the formula for this procedure is thus
\begin{equation} \label{eq:extrap_minus_upper} %see LBXXX p.117
\tilde\alpha_{\text{c}^-,\,\infty}^\text{ub}=n_\text{max}\tilde\alpha_{\text{c}^-,\,n_\text{max}}-\left(n_\text{max}-1\right)\tilde\alpha_{\text{c}^-,\,n_\text{max}-1}.
\end{equation}

In this way we find, to four significant figures:
$\tilde\alpha^\text{\,sc}_{\text{c}^+,\,\infty}=0.1868^{\left(0.1876\right)}_{\left(0.1860\right)}$,
$\tilde\alpha^\text{\,sc}_{\text{c}^-,\,\infty}=-0.1032^{\left(-0.1026\right)}_{\left(-0.1039\right)}$,
$\tilde\alpha^\text{\,bcc}_{\text{c}^+,\,\infty}=0.2388^{\left(0.2395\right)}_{\left(0.2380\right)}$,
$\tilde\alpha^\text{\,bcc}_{\text{c}^-,\,\infty}=-0.1194^{\left(-0.1188\right)}_{\left(-0.1199\right)}$,
$\tilde\alpha^\text{\,fcc}_{\text{c}^+,\,\infty}=0.2388^{\left(0.2396\right)}_{\left(0.2379\right)}$, and
$\tilde\alpha^\text{\,fcc}_{\text{c}^-,\,\infty}=-0.1194^{\left(-0.1188\right)}_{\left(-0.1199\right)}$, where superscripts denote upper bounds and subscripts denote lower bounds.  (Here, we have rounded away from zero for the upper bounds of $\tilde\alpha_{\text{c}^+,\,\infty}$ and the lower bounds of $\tilde\alpha_{\text{c}^-,\,\infty}$, and rounded towards zero for the lower bounds of $\tilde\alpha_{\text{c}^+,\,\infty}$ and the upper bounds of $\tilde\alpha_{\text{c}^-,\,\infty}$, to preserve, when rounded, the integrity of the values as bounds.)  It is seen that, in all cases, the upper and lower bounds lie within $\pm 0.6\,\%$ of the best-estimate values, hence we may summarize that the values of $\tilde\alpha_{\text{c}^\pm,\,\infty}$ obtained by the above method (with $n_\text{max}=27$) are accurate to within $\pm 0.6\,\%$ for each crystal structure, according to this way of assessing the accuracy.  The values for the bcc and fcc cases are consistent with, and suggestive of, the true model values of $\tilde\alpha^\text{\,bcc}_{\text{c}^+,\,\infty}$ and $\tilde\alpha^\text{\,fcc}_{\text{c}^+,\,\infty}$ being exactly $\frac{3}{4\pi}$ (as assumed previously), and the true model values of $\tilde\alpha^\text{\,bcc}_{\text{c}^-,\,\infty}$ and $\tilde\alpha^\text{\,fcc}_{\text{c}^-,\,\infty}$ being exactly $-\frac{3}{8\pi}$ (as deduced previously).

Inserting the best-estimate, upper bound, and lower bound values of $\tilde\alpha_{\text{c}^\pm,\,\infty}$ into Eq.~(\ref*{eq:cm_reduced}), we find:
$\chi^{(0)\,\text{sc}}_{\text{c}^+}=10.80^{\left(10.99\right)}_{\left(10.60\right)}$,
$\chi^{(0)\,\text{sc}}_{\text{c}^-}=-0.9057^{\left(-0.9020\right)}_{\left(-0.9093\right)}$,
$\chi^{(0)\,\text{bcc}}_{\text{c}^-}=-1.0001^{\left(-0.9971\right)}_{\left(-1.0030\right)}$,
and $\chi^{(0)\,\text{fcc}}_{\text{c}^-}=-1.0001^{\left(-0.9972\right)}_{\left(-1.0029\right)}$.
Again, the best-estimate values, and bounds, for $\chi^{(0)\,\text{bcc}}_{\text{c}^-}$ and $\chi^{(0)\,\text{fcc}}_{\text{c}^-}$ are consistent with, and suggestive of, true model values of exactly $-1$.  As percentages, the accuracy of $\chi^{(0)\,\text{sc}}_{\text{c}^+}$ is within $\pm 1.8\,\%$ and the accuracies of $\chi^{(0)\,\text{sc}}_{\text{c}^-}$, $\chi^{(0)\,\text{bcc}}_{\text{c}^-}$, and $\chi^{(0)\,\text{fcc}}_{\text{c}^-}$ are all within $\pm 0.5\,\%$.

This way of assessing the accuracy has the advantage, compared to that presented in \S\ref*{sec:acc_comp}, that it does not require any of the values of $\tilde\alpha_{\text{c}^\pm,\,\infty}$ to be assumed \textit{a priori} for `calibration', and it does not require us to assume that the accuracy of the values of  $\tilde\alpha_{\text{c}^\pm,\,\infty}$ are similar for each crystal structure (though it ends up confirming that this is the case).  In this regard, it may be considered somewhat more rigorous.  On the other hand, it leads to less-stringent specifications of the accuracy.

\subsection{Study to rule out sample-shape dependence}

If our method is valid, it must be the case that the results obtained for $\tilde\alpha_{\text{c}^\pm,\,\infty}$ and $\chi^{(0)}_{\text{c}^\pm}$ are independent of the overall shape of the crystals (assuming the crystals are macroscopic in all three dimensions, i.e., not 2D or 1D arrays).  We have carried out a study to confirm that this is the case \cite{dutta_castles_unpublished}, which may be summarized as follows.

In addition to the rhombohedral samples reported above, for each of the sc, bcc, and fcc cases, the following, alternatively-shaped, crystals were studied:
\begin{itemize}
	\item `Parallelepiped slabs', formed from entities at the sets of position vectors 
	\[
	S_n^\text{\,slab}=\left\{\tilde{\mathbf{r}}^{(u_1,u_2,u_3)}:u_1,u_2=0...n^2-1; u_3=0..n-1\right\}.
	\]
	(More specifically, slab-like square cuboids in the sc case.)
	\item `Parallelepiped needles', formed from entities at the sets of position vectors
	\[
	S_n^\text{\,needle}=\left\{\tilde{\mathbf{r}}^{(u_1,u_2,u_3)}:u_1,u_2=0...n-1; u_3=0..n^2-1\right\}.
	\]
	(More specifically, needle-like square cuboids in the sc case.)
	\item Spheres, formed from entities at the sets of position vectors
	\[S_n^\text{\,sphere}=\left\{\tilde{\mathbf{r}}^{(u_1,u_2,u_3)}:u_1,u_2,u_3\in\mathbb{Z}; \left|\tilde{\mathbf{r}}^{(u_1,u_2,u_3)}\right|\leq n\right\}.
	\]
\end{itemize}
In each case, if the values of $\tilde\alpha_{\text{c}^+,\,n}$ and $\tilde\alpha_{\text{c}^-,\,n}$ for set $S_n$ are calculated for a range of $n$, the limits $\tilde\alpha_{\text{c}^\pm,\,\infty}=\lim_{n\rightarrow\infty}\tilde\alpha_{\text{c}^\pm,\,n}$ are determined by extrapolation, and the accuracy is determined by one or other of the methods above---i.e., if the same process is carried out for these alternatively-shaped crystals as was carried out for the rhombohedral crystals above---then the same values of $\tilde\alpha^{\,\text{sc}}_{\text{c}^\pm,\,\infty}$, $\tilde\alpha^{\,\text{fcc}}_{\text{c}^\pm,\,\infty}$, and $\tilde\alpha^{\,\text{bcc}}_{\text{c}^\pm,\,\infty}$, and hence the same values of $\chi^{(0)\,\text{sc}}_{\text{c}^\pm}$, $\chi^{(0)\,\text{fcc}}_{\text{c}^\pm}$, and $\chi^{(0)\,\text{bcc}}_{\text{c}^\pm}$ are found, within the accuracy of the method (i.e., the same results as already reported in Table~\ref*{tab:data}).

Of course, this approach does not rule out the possibility that some other overall shape of crystals that we have not checked explicitly---say, ellipsoidal---may somehow give different values, but we nevertheless consider it appropriate to reasonably conclude that the method appears to provide results that are independent of the sample shape.

\subsection{An alternative methodology (yields the same results)} \label{sec:alternative_method}

We may refer to the method employed hitherto in this paper as the `finite crystal method' since the macroscopic limit is considered by extrapolating the results for finite crystals of increasing size.  An alternative methodology involves assuming an infinite crystal from the outset, which we may refer to as the `infinite crystal method'.  Both methods were considered and applied by Allen to investigate the sc case with positive polarizability in Ref.~\cite{allen_2004}.  Each method has certain advantages and disadvantages, but we believe that, overall, the finite crystal method provides the most direct and rigorous route to determine the values of $\tilde\alpha_{\text{c}^\pm}$ with well-defined accuracy in the macroscopic limit, and hence it is the method we reported in detail above.  However, we have also carried out a detailed analysis of the infinite crystal method as applied to sc, bcc, and fcc crystals with emphasis on the negative static polarizability case \cite{dutta_castles_unpublished}; in all cases, we find excellent agreement with the above-stated results, providing a useful cross-check and validation.

One favorable feature of the infinite crystal method is that, once the location in reciprocal space of the extremal eigenvalues has been established for a given crystal structure, the values of $\tilde\alpha_{\text{c}^\pm}$ (which are, necessarily, the macroscopic values) may be expressed as infinite lattice sums.  For example, in the sc case we may write
\begin{equation} \label{eq:lattice_sum_plus}
\tilde{\alpha}^\text{\,sc}_{\text{c}^+}=\left[\sum_{\mathsf{u}'}\frac{(-1)^{u_2+u_3}\left(2u_1^2-u_2^2-u_3^2\right)}{\left(u_1^2+u_2^2+u_3^2\right)^{5/2}}\right]^{-1},
\end{equation}
and
\begin{equation} \label{eq:lattice_sum_minus}
\tilde{\alpha}^\text{\,sc}_{\text{c}^-}=\left[\sum_{\mathsf{u}'}\frac{(-1)^{u_1}\left(2u_1^2-u_2^2-u_3^2\right)}{\left(u_1^2+u_2^2+u_3^2\right)^{5/2}}\right]^{-1},
\end{equation}
where $\sum_{\mathsf{u}'}$ denotes the sum over all triples of integers $\mathsf{u}=\left(u_1,u_2,u_3\right)$ except $\left(0,0,0\right)$ (corresponding to the interaction of a given dipole with all \textit{other} dipoles in the infinite crystal).  We are not aware of any closed form expressions for these two particular sums, but they are absolutely convergent and may be readily approximated either via brute force summation with $u_1, u_2, u_3=-N_\text{sum}...N_\text{sum}$ for some large $N_\text{sum}$, by evaluating the sum for a range of $N_\text{sum}$ and extrapolating $N_\text{sum}\rightarrow\infty$, or by using the Ewald summation method.  To three significant figures, the values of the sums are $\tilde{\alpha}^\text{\,sc}_{\text{c}^+}=0.187$ and $\tilde{\alpha}^\text{\,sc}_{\text{c}^-}=-0.103$, which agree with the results found previously for the finite crystal method recorded in Table~\ref*{tab:data}.

The reciprocal $\lambda_\text{max}^\text{sc}=1/\tilde{\alpha}^\text{\,sc}_{\text{c}^+}=5.35$ was considered explicitly by Allen and is identified as the maximum point of the graph in Fig.~3 of Ref.~\cite{allen_2004}.  The lower critical values were not of interest to Allen, but we may see that the reciprocal $\lambda_\text{min}^\text{sc}=1/\tilde{\alpha}^\text{\,sc}_{\text{c}^-}=-9.69$ is the minimum point of the graph in Fig.~3 of Ref.~\cite{allen_2004}.  (We are not aware of any general theorem that states the extremal eigenvalues must lie along the edges of the irreducible Brillouin zone, as Allen appears to assume, but our work \cite{dutta_castles_unpublished} indicates that this is indeed true for the sc, bcc, and fcc cases we have considered.)

The lattice sums of Eqs.~(\ref*{eq:lattice_sum_plus}) and (\ref*{eq:lattice_sum_minus}) appear also in other physical systems and have, long ago, been evaluated in other contexts.  For example, they appear in the related problem of the \textit{preferred orientation} of arrays of electric or magnetic dipoles with \textit{fixed magnitudes}: in this context, the value of what we refer to as $\lambda_\text{max}^\text{sc}$, for example, appears in Ref.~\cite{luttinger_1946} (denoted as ``$f_5$'' and listed in Table II of that reference), and in Ref.~\cite{sauer_1940} (the value of $-2.7$ quoted for the ``Antiparallel `A' '' structure with ``Lattice structure $a$'' in Table I of Ref.~\cite{sauer_1940} is essentially $-\lambda_{\text{max}}^\text{sc}/2$).  The value of what we refer to as $\lambda_\text{min}^\text{sc}$ appears also in Ref.~\cite{luttinger_1946} (denoted as ``$f_2$'' and listed in Table II of Ref.~\cite{luttinger_1946}).

\section{Conclusions}

We have found that static electric susceptibility values are possible, according to the model, for non-equilibrium cubic crystals (with one dipolarizable entity per primitive cell) in the following intervals:
	\begin{align*}
	-0.906 &\lesssim \chi^{(0)\,\text{sc}} \lesssim 10.8, \\
	-1.00&\lesssim \chi^{(0)\,\text{bcc}}, \\
	-1.00&\lesssim \chi^{(0)\,\text{fcc}}.
	\end{align*}
The analysis of \S\ref*{sec:accuracy} indicates that the numerically-determined endpoints for the intervals are almost certainly accurate to the three significant figures stated and the lower endpoints for bcc and fcc are consistent with, and highly suggestive of, the true model values being exactly $-1$ in both cases.

All values of $\chi^{(0)}$ within the intervals can, according to the model, be obtained via an appropriate value of $\tilde\alpha$.  The value of $\tilde\alpha$ necessary to produce a given value of $\chi^{(0)}$ may be obtained from the Clausius-Mossotti equation [by inverting the version of the equation presented in Eq.~(\ref*{eq:cm_reduced})].  Although we assume that $\tilde\alpha$ may, in principle, take any (real) value, hypothetical values $\chi^{(0)\,\text{sc}} \gtrsim 10.8$, $\chi^{(0)\,\text{sc}} \lesssim -0.906$, $\chi^{(0)\,\text{bcc}} \lesssim -1.00$, and $\chi^{(0)\,\text{fcc}} \lesssim -1.00$ are nevertheless impossible, within the model, because there is no value of $\tilde\alpha$ that would result in a \textit{stable} material with such a value of $\chi^{(0)}$.  We therefore refer to the intervals above as the \textit{permissible intervals} of $\chi^{(0)}$ and the lower endpoints of the above intervals as the lower \textit{permissible bounds} of $\chi^{(0)}$ for the given crystal structure \footnote{Note, however, that we make no comment on whether the system is stable for values of $\tilde\alpha$ exactly equal to $\tilde\alpha_{\text{c}^-,\,\infty}$ and, therefore, whether values of $\chi^{(0)}$ exactly equal to $\chi^{(0)}_{\text{c}^-}$ are possible.}.  The lower permissible bounds are unrelated to the horizontal asymptote of the Clausius-Mossotti curve (as it is presented in Fig.~\ref*{fig:cm}) and they require a more-sophisticated approach than the standard Clausius-Mossotti analysis to derive.

Our initial motivation for carrying out the work presented herein was a concern that mutual interactions between entities with $\alpha<0$ might, collectively, wipe out the possibility of $\chi^{(0)}<0$ materials in the macroscopic limit: that is, we were concerned that, in the limit of large particle numbers, the lower critical value of $\tilde\alpha$ would converge to zero and, accordingly, the lower permissible bound of $\chi^{(0)}$ would be zero in all cases.  (The experiments reported in Ref.~\cite{castles_2020} were not sophisticated enough to rule out this possibility.)  Perhaps most importantly, then, we have seen that this does not happen; the lower permissible bounds of $\chi^{(0)}$ are negative definite in each case.  Therefore, we conclude that negative static electric susceptibility values are indeed possible in non-equilibrium crystals, according to the model. (We note that this conclusion is very different from that of the circuit theory argument, which also showed that values $\chi^{(0)}<-1$ are impossible whether the material is in equilibrium or not \textit{but did not imply that any values in the interval} $-1<\chi^{(0)}<0$ \textit{are necessarily possible}.)

We see that the lower permissible bound depends on the crystal structure: it is different for sc than for bcc and fcc.  Thus, the question of the lower permissible bound for a given crystal structure appears to be, in general, a highly non-trivial question which can probably only be addressed definitively by the type of methods considered herein.  In particular, we can assert that there is no structure-independent `short cut' derivation of the lower permissible bound.

Although the value $\chi^{(0)\,\text{sc}}_{\text{c}^-}=-0.906$ differs by only $\approx10\,\%$ from the values $\chi^{(0)\,\text{bcc}}_{\text{c}^-}=\chi^{(0)\,\text{fcc}}_{\text{c}^-}=-1.00$, and the assumption of point-like dipolarizable entities means that the results of the study may be inaccurate in relation to any real system, our results suggest that, if one wishes to create an isotropic material with the lowest possible value of $\chi^{(0)}$, a bcc or fcc structure, as opposed to a sc structure, is likely to be preferable.

It is clear that the case of $\chi^{(0)}<0$ is very different from the usual case of $\chi^{(0)}>0$.  Whereas the permissible value of $\chi^{(0)}$ may be unbounded above for certain crystal structures (e.g., bcc and fcc with one entity per primitive cell), it is always bounded below by the value $-1$ according to the circuit theory argument, and, as we have seen, may be even more limited in its negative extent in certain crystal structures (e.g., sc with one entity per primitive cell).  Nevertheless, in finding that the permissible interval may extend all the way down to the circuit theory limit of $-1$ for certain condensed media, we have confirmed that condensed media are, indeed, capable of exhibiting $\chi^{(0)}<0$ values with magnitudes that are $\sim 10^3$ times greater than those proposed initially in gaseous systems, even when mutual interactions between the dipolarizable entities are taken into account.  We believe that this increase is likely to be sufficient to enable the remarkable potentialities of $\chi^{(0)}<0$ materials---such as new forms of charged particle traps---to become practically feasible.

\section*{Acknowledgments}
%\medskip

This work was funded by the Engineering and Physical Sciences Research Council UK (Grant No. EP/R035393/1).

%\bibliography{C:/Users/Flynn/MyDriv\string~1/biblibrary_FC_GDrive/references_FC}

%

\end{document}